\documentclass[iop]{emulateapj}
\usepackage[english]{babel}
\usepackage{amsmath}
\usepackage{graphicx}
\usepackage{wasysym}
\usepackage{soul}
\usepackage[dvipsnames]{xcolor}
\usepackage{parskip}
\usepackage{wrapfig}
\usepackage{url}
\usepackage{float}
\newcommand{\rstar}{$2.3\pm0.2~\mathrm{R_\odot}$}
\newcommand{\rotperiod}{$8.81^{+1.9}_{-1.8}~\mathrm{hr}$}
\newcommand{\obliq}{$69^{\circ} \pm 3^\circ$}
\newcommand{\oblate}{$0.19^{+0.04}_{-0.03}$}
\newcommand{\limbdarkone}{0.56}
\newcommand{\limbdarktwo}{-0.15}
\newcommand{\flux}{$1.000009 \pm 3*10^{-6}$}
\newcommand{\rpone}{$0.45 \pm 0.03~\mathrm{R_{jup}}$}
\newcommand{\tcenterone}{$34960800 \pm 400~\mathrm{s}$}
\newcommand{\periodone}{84.69}
\newcommand{\incone}{$89.340^\circ \pm 0.05^\circ$}
\newcommand{\bone}{$0.61^{+0.08}_{-0.07}$}
\newcommand{\eone}{$0.056 \pm 0.019$}
\newcommand{\ascone}{$-32^{\circ} \pm 11^\circ$}
\newcommand{\alignone}{$72^{\circ} \pm 3^\circ$}
\newcommand{\rptwo}{$0.43 \pm 0.05~\mathrm{R_{jup}}$}
\newcommand{\tcentertwo}{$25041400 \pm 700~\mathrm{s}$}
\newcommand{\periodtwo}{207.58}
\newcommand{\inctwo}{$90.64^\circ \pm 0.06^\circ$}
\newcommand{\btwo}{$-0.57^{+0.12}_{-0.14}$}
\newcommand{\etwo}{$0.50 \pm 0.09$}
\newcommand{\asctwo}{$-32^\circ \pm 40^\circ$}
\newcommand{\aligntwo}{$73^{\circ+11}_{~-5}$}

\newcommand{\nomech}{ten~}

\newcommand{\noposs}{five~}

\newcommand{\noimposs}{five~}

\begin{document}
\pagestyle{myheadings}
\title{Spin-Orbit Misalignment of Two-Planet-System KOI-89 Via Gravity Darkening}

\author{John P. Ahlers, Jason W. Barnes} 
\affil{Physics Department, University of Idaho, Moscow, ID 83844}
\author{Rory Barnes}
\affil{Astronomy Department, University of Washington, Seattle, WA 98195}

\begin{abstract}
We constrain the true spin-orbit alignment of the KOI-89 system by numerically fitting the two \emph{Kepler} photometric lightcurves produced by transiting planets KOI-89.01 and KOI-89.02. The two planets have periods of \periodone~days and \periodtwo~days, respectively. We find that the two bodies are low-density giant planets with radii \rpone~and \rptwo~and spin-orbit misalignments \alignone~and \aligntwo, respectively. Via dynamic stability tests we demonstrate the general trend of higher system stability with the two planets close to mutual alignment and estimate their coalignment angle to $20^\circ \pm 20^\circ$ -- i.e. the planets are misaligned with the star but may be aligned with each other. From these results, we limit KOI-89's misalignment mechanisms to star-disk-binary interactions, disk warping via planet-disk interactions, planet-planet scattering, Kozai resonance, or internal gravity waves. 
\end{abstract}


\section{Introduction} \label{sec:intro}
\footnotetext[1]{ http://www.physics.mcmaster.ca/\texttt{\char`\~}rheller/ provides a list of spin-orbit misaligned planets.}

Recent studies show that exoplanetary systems around early-type stars display a high diversity in their fundamental characteristics with varied planet radii, planet densities, periods, eccentricities, and inclinations.  \citep{brandt2014analysis, borucki2012kepler, howard2013observed}. Exoplanet orbits have notably varied spin-orbit alignments with highly misaligned and even retrograde orbit geometries \citep{2014A&A...569A..65B, 2012ApJ...757...18A, 2011AJ....141...63W}. At the time of this work, most known misaligned systems are short-period; only HD80806b \citep{naef2001hd}, Upsilon Andromedae \citep{barnes2015upsilon}, and Kepler 56 \citep{huber2013stellar} have periods $\geq 10~\mathrm{days}$.\footnotemark[1] This work adds to the list of long-period spin-orbit misaligned planets KOI-89.01 and KOI-89.02. 

There are several methods for determining various aspects of a system's alignment, including gravity-darkening \citep[J.][]{2011ApJS..197...10B,ahlers2014}, the Rossiter-McLaughlin effect \citep{ohta2005rossiter}, Doppler tomography \citep{gandolfi2012doppler}, asteroseismology \citep{chaplin2013asteroseismic,2014ApJ...782...14V}, photometric amplitude distribution \citep{2015ASPC..496..167M}, and stroboscopic starspots \citep{desert2011hot,0004-637X-756-1-66}. We applied the gravity-darkening method first suggested by J. \citet{2009ApJ...705..683B} and later applied to Kepler Object of Interest (KOI) 13 \citep[J.][]{2011ApJS..197...10B} and KOI-2138 \citep[J.][]{barnes2015probable}. This method constrains both the star's polar tilt toward/away from the observer (stellar obliquity), and the planet's misalignment angle as seen relative to the observer (sky-projected alignment).

The gravity-darkening effect, first predicted by \citet{von1924radiative}, results in a pole-to-equator gradient in stellar luminosity driven by rotation. As an object transits a gravity-darkened star, it can move across areas of unequal brightness; this luminosity gradient can affect the lightcurve in various ways, depending on its transit geometry \citep[J.][]{2009ApJ...705..683B}. J. \citet{2011ApJS..197...10B}, J. \citet{barnes2013measurement}, \citet{zhou2013highly}, and  \citet{ahlers2014} all showed that the asymmetry in such lightcurves (or lack thereof) can be utilized to constrain the spin-orbit alignment of a transiting system.

The causes of frequent misalignment around fast rotators are still under investigation. The underlying issue is that planets probably do not form with initially misaligned orbits -- their angular momenta must be conserved with the stellar nursery they formed in. In this case, such planets must migrate to their misaligned positions. KOI-89 is one such system that does not conform to the traditional nebular hypothesis.

There are several ideas for processes that might create spin-orbit misalignment. \citet{lai2011evolution} and \citet{spalding2014primordial} demonstrated that magnetic torques can push the stellar spin axis away from the circumstellar disk's angular momentum vector over very long timescales. This would specifically explain spin-orbit migration in very young systems with late-type stars, where stellar magnetic fields are strongest.  \citet{2012ApJ...758L...6R} showed that internal gravity waves can produce angular momentum transport between the convective interior and radiative exterior of early-type stars that turn the stellar spin axis away from the system's invariant plane.

There are also several ideas that explain how spin-orbit migration might develop via more dynamic means. \citet{libert2009kozai} discussed Kozai resonance in a 2-planet system and its effects on mutual inclination. This is almost certainly the origin of misalignment for HD 80806b \citep{naef2001hd}. \citet{chatterjee2008dynamical}, \citet{ford2005planet}, \citet{raymond2008mean}, and \citet{nagasawa2008formation} all demonstrated how planet-planet scattering can drive misalignment in a multiplanet system. \citet{levison1998modeling} showed that planet-embryo collisions during planet formation can lead to high mutual inclination. \citet{terquem2013effects}, \citet{teyssandier2012orbital}, and \citet{batygin2012primordial} analyze gravitational disk-warping events that lead to misalignment. 

\citet{winn2010hot} showed a correlation between hot stars ($T_{\mathrm{eff}}\gtrsim 6250$) and misalignment. \citet{batygin2013magnetic} showed an interdependence between stellar rotation rates and spin-orbit misalignments. These works imply that a large number of planets orbiting early-type stars are commonly misaligned.

\citet{huber2013stellar} employed asteroseismology to measure the the stellar obliquity of multiplanet system Kepler 56, and showed that spin-orbit misalignment is possible in multiplanet systems with low-mass, long-period planets. \citet{2014PASJ...66...94B} found mild misalignment in Kepler-25 via a joint analysis of asteroseismology, lightcurve analysis, and the Rossiter-Mclaughlin effect.

This work provides another example of a long-period multiplanet system with significant misalignment: KOI-89. In \S\ref{sec:observations}, we outline our data preparation process and list previously known system characteristics. In \S\ref{sec:model}, we introduce new techniques to the J. \citet{2011ApJS..197...10B} fitting method. In \S\ref{sec:results}, we show our best-fit parameters and constraints on misalignment. We test KOI-89's dynamic stability and constrain the coalignment angle between the two orbits in \S\ref{sec:stability}. In \S\ref{sec:discussion}, we discuss possible formation and migration mechanisms for the KOI-89 system, as well as test the dynamic stability of the system in order to constrain the planets' mutual alignment.

\section{Observations} \label{sec:observations}

\subsection{Data Preparation}\label{sec:preparation}

The Mikulski Archive for Space Telescopes (MAST) Kepler Input Catalog (KIC) provided the \emph{Kepler} photometry that we analyze for the KOI-89 system. We employ each of the 16 available quarters of KIC data, combining them into a single dataset. We choose to only incorporate long cadence data (30-minute integrations) because both planets have transit durations of over 12 hours. Therefore ingress and egress are well-sampled by the 30 minute time cadence, and inclusion of short (1-minute) cadence data would not provide additional constraints. 

\begin{figure}[tbhp]
\epsscale{1}
\plotone{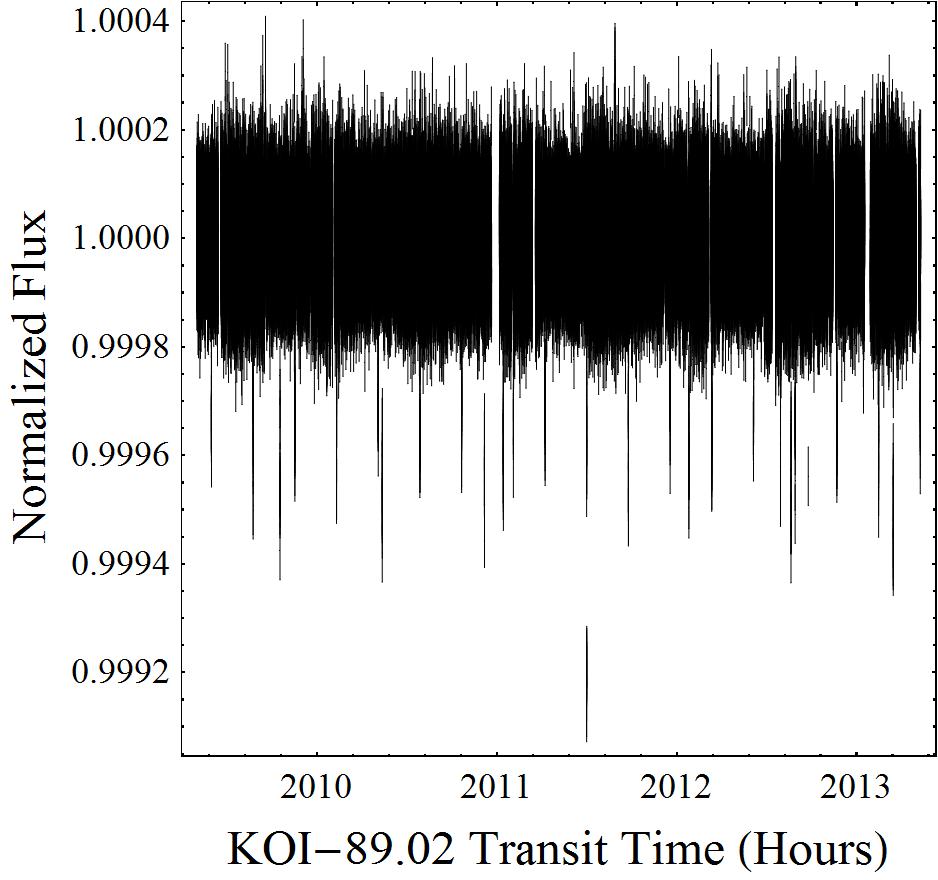}
\caption{\footnotesize \emph{Kepler} photometry of the KOI-89 system. The vertical length of the data points represent their uncertainties. Transits of two planets are visible, displaying periodicities of \periodone~days and \periodtwo~days, respectively. In mid-2011 (arrow), both planets transited simultaneously.
}
\label{fig:alldata}
\end{figure}

After concatenating all available long cadence photometry, we apply a median box filter of 44 hours (three times KOI-89.01's transit duration) to reduce long-term astrophysical and instrumental variability. Figure \ref{fig:alldata} displays the filtered time-series. We then identify which transits correspond to which transiting body based on their KIC orbital periods, and separate them accordingly into individual datasets. 

%{\bf FIX ACCORDING TO HOW SHIFTED FITS GO} We fold all individual KOI-89.01 and KOI-89.02 lightcurves on top of each other, excluding the double transit. \citet{rowe2014validation} measured transit timing variations (TTV) $TTV$ for both planets; during the folding process, we account for these variations by manually adjusting each transit against its respective $TTV_n$. However, \citet{rowe2014validation} does not include KOI-89's $TTV_n$ for \emph{Kepler} quarters 13-16, so we only adjust transits with known $TTV$ values.  While the $TTV_n$ are significant ($\sim 20$~min), we find that they negligibly affected our folding process, and were effectively removed during data preparation. Therefore, we fold in all available \emph{Kepler} photometry, including transits without known $TTV$. 

We adjust the center-of-transit times of each transit lightcurve according to their measured transit timing variations ($TTV$) \citep{rowe2014validation}. We perform this adjustment for each measured TTV in \citet{rowe2014validation}, including thirteen transits for KOI-89.01 and five transits for KOI-89.02. These transits exclude the double transit identified in Figure \ref{fig:alldata}. 

With all $TTV$ accounted for and the individual transits evenly separated by \periodone~days and \periodtwo~days, respectively, we fold all KOI-89.01 transits on top of the epoch \tcenterone~transit and fold all KOI-89.02 transits on top of the epoch \tcentertwo~transit. We then combine the two resulting lightcurves back in a single dataset. With the \emph{Kepler} photometry represented by a single lightcurve with two transit events, we bin the data at 15 minutes to improve the computation time of our fit. We determine the error bars of the binned data from the standard deviation of the flux values in the bin.

\subsection{Previously Measured KOI-89 Parameters}

\begin{figure*}[th]
\epsscale{1}
\includegraphics[trim=0 295 60 70, clip, width=8in]{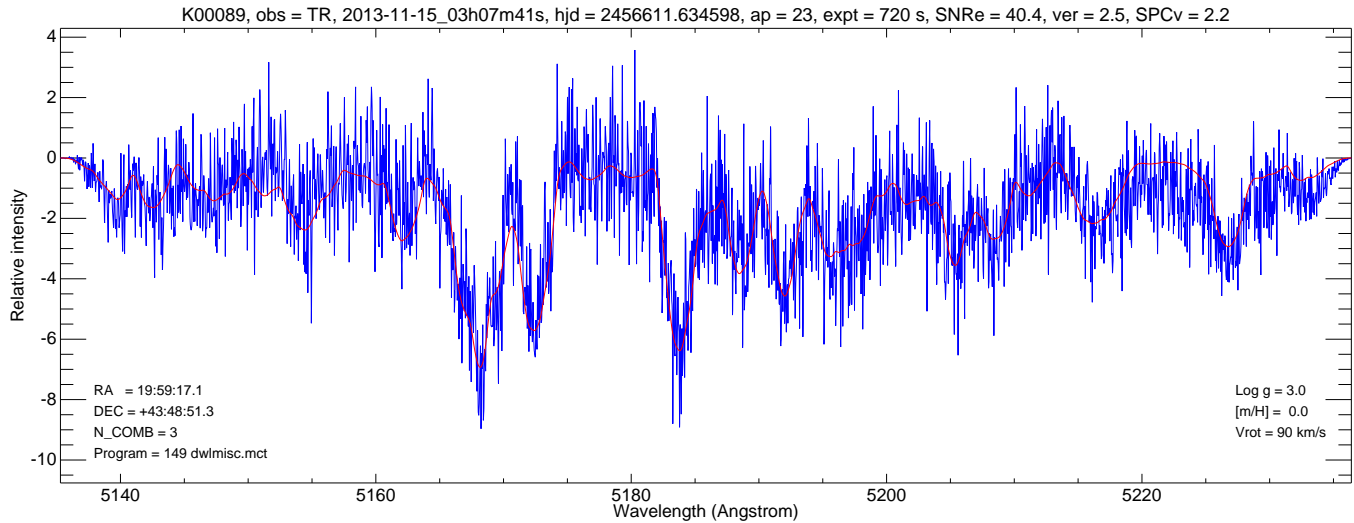}
\caption{\footnotesize Spectroscopic determination of $v\sin(i)$ for the KOI-89 system, measured with the Tillinghast Reflector Echelle Spectrograph on the 1.5 m telescope at the Whipple observatory. These data were provided by David Latham of the \emph{Kepler} Science Team and retrieved from the Community Follow-up Observing Program (CFOP).}
\label{fig:vsini}
\end{figure*}

We build our work upon previous research of the KOI-89 system. We obtain the KOI-89 stellar mass, stellar temperature, and transit periods from the Community Follow-up Observation Program (CFOP). Figure \ref{fig:vsini} shows the spectroscopic determination of $v\sin(i)$. We list these and other relevant system parameters in Table \ref{table:prevdata}. 

\renewcommand{\arraystretch}{1.3}
\begin{table}[hbtp]
\centering
\begin{tabular}{l r}
\hline
{\bf Parameter} & {\bf Values} \\ \hline
\hline

$P_1$ & \periodone ~$\mathrm{days}$ \\
$P_2$ & \periodtwo ~$\mathrm{days}$ \\
$T_{\mathrm{eff}}$ & $7717 \pm 225~\mathrm{K}$\\
$M_\star$ & $1.965 \pm 0.256~\mathrm{M_\odot}$ \\
$v\sin(i)$ & $90~\mathrm{km/s}$ \\
$V_{\mathrm{mag}}$ & $11.731$ \\
KOI-89.01 SNR & 93.8 \\
KOI-89.02 SNR & 68.1 \\
 \\ 
\end{tabular}
\caption{\footnotesize Previously measured parameters of the KOI-89 system. We incorporated all parameters as assumed values when fitting the KOI-89 lightcurve.  
}
\label{table:prevdata}
\end{table}

\citet{rowe2014validation} confirmed 715 new systems -- including KOI-89 -- via multiplicity. The two planets have a period ratio near the 5:2 mean-motion resonance (2.45). Follow-up observations of these phenomena could confirm/deny the existence of additional orbiting bodies, and could further constrain this system's formation and evolution.

\section{Model} \label{sec:model}

We update the {\tt transitfitter} program \citep[J.][]{2009ApJ...705..683B} to fit multiple-planet transiting systems. The Levenburg-Marqhardt $\chi^2$ minimization technique remains the fitting method, but now {\tt transitfitter} can constrain the orbital parameters of additional transiting bodies. These extra parameters cause an increased sensitivity to the photometric signal-to-noise ratio, so $\chi^2$ minimization must be approached with additional caution.

The individual parameters of additional planets have the same degeneracies as a single-planet fit. There is a degeneracy between eccentricity and stellar radius, for instance: an eccentric orbit can have the same transit duration as a circular orbit around a smaller star. Also, fitting transit lightcurves in isolation cannot determine stellar mass, so we apply an assumed stellar mass from CFOP and fit the eccentricity around it. A transit around a fast-rotator has degenerate limb-darkening and gravity-darkening effects in the case of high stellar obliquity. We discuss this degeneracy further in \S\ref{sec:results}.
 
\begin{figure}[tbhp]
\centering
\begin{tabular}{c|c}
\includegraphics[scale=0.88]{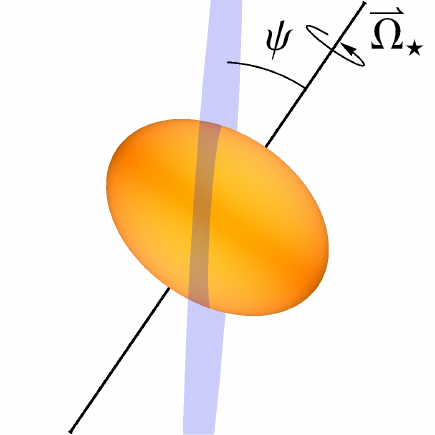} & \includegraphics[scale=0.88]{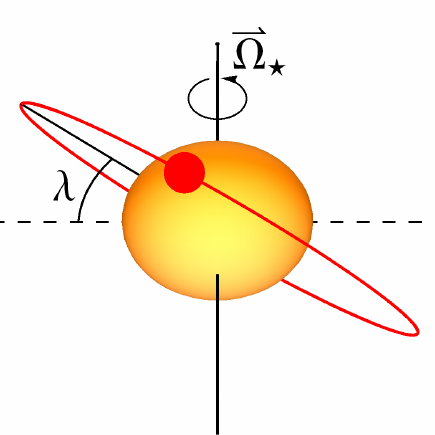} \\
\hline \\
\includegraphics[scale=0.88]{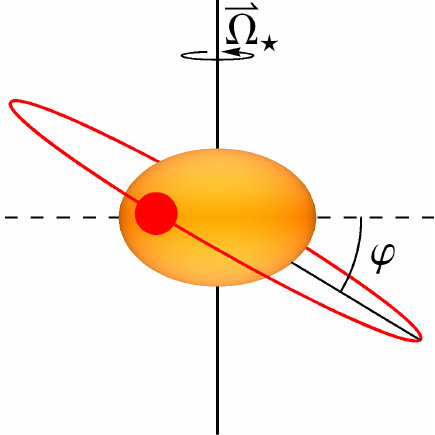} & \includegraphics[scale=0.88]{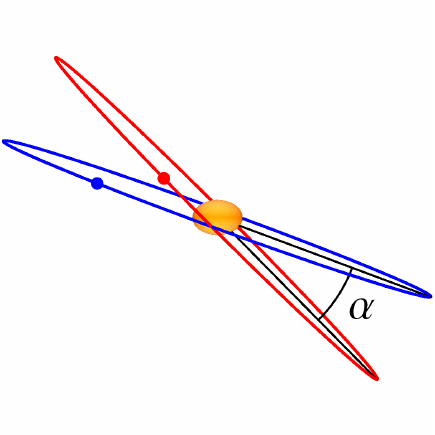}
\end{tabular}
\caption{\footnotesize Top left: The stellar obliquity $\psi$ is defined as the axial tilt toward/away the plane of the sky. Top right: the sky-projected alignment $\lambda$ is the misalignment angle seen from \emph{Kepler's} point of view. Bottom left: the spin-orbit alignment $\varphi$ is the angle between the plane of an orbit and the star's equatorial plane. Bottom right: the coalignment angle $\alpha$ is the angle between two orbit planes.}
\label{fig:definitions}
\end{figure}

Fitting for both planets simultaneously rather than fitting each lightcurve separately forces the stellar radius $R_\star$ and the stellar obliquity $\psi$ (shown in Figure \ref{fig:definitions}) to agree for both planets, which has two advantages. First, a simultaneous two-planet fit requires two fewer parameters to be fit, increasing the overall accuracy and decreasing computation time of the other parameters. Second, it applies lightcurve geometries of both planets toward the constraint of $R_{\star}$ and $\psi$, producing a coherent physical system. In our fitting model, $\psi$ is particularly sensitive to the lightcurve geometry; using multiple lightcurves simultaneously allows $\psi$ to be constrained by more data points, decreasing its uncertainty. 

\section{Results} \label{sec:results}

\subsection{Lightcurve Fits} \label{sec:fits}

\begin{figure*}[tbhp]
\begin{tabular}{r l}
\vspace*{-1cm} \\
\includegraphics[width=0.47\textwidth]{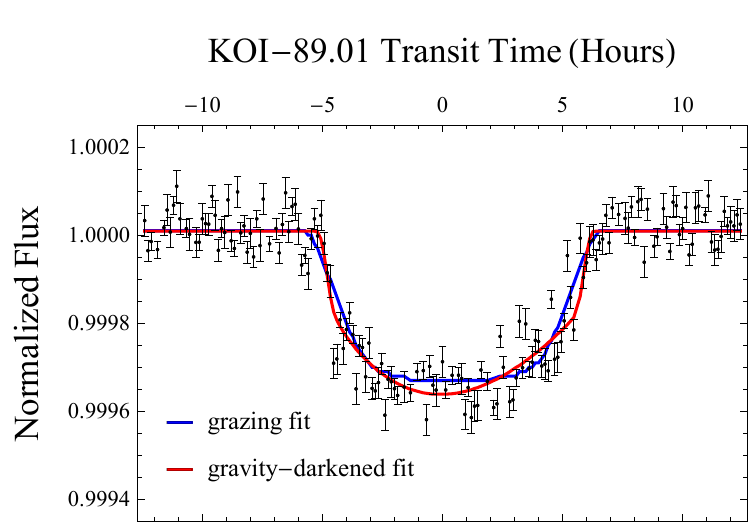} & \hspace*{-0.37cm}\includegraphics[width=0.47\textwidth]{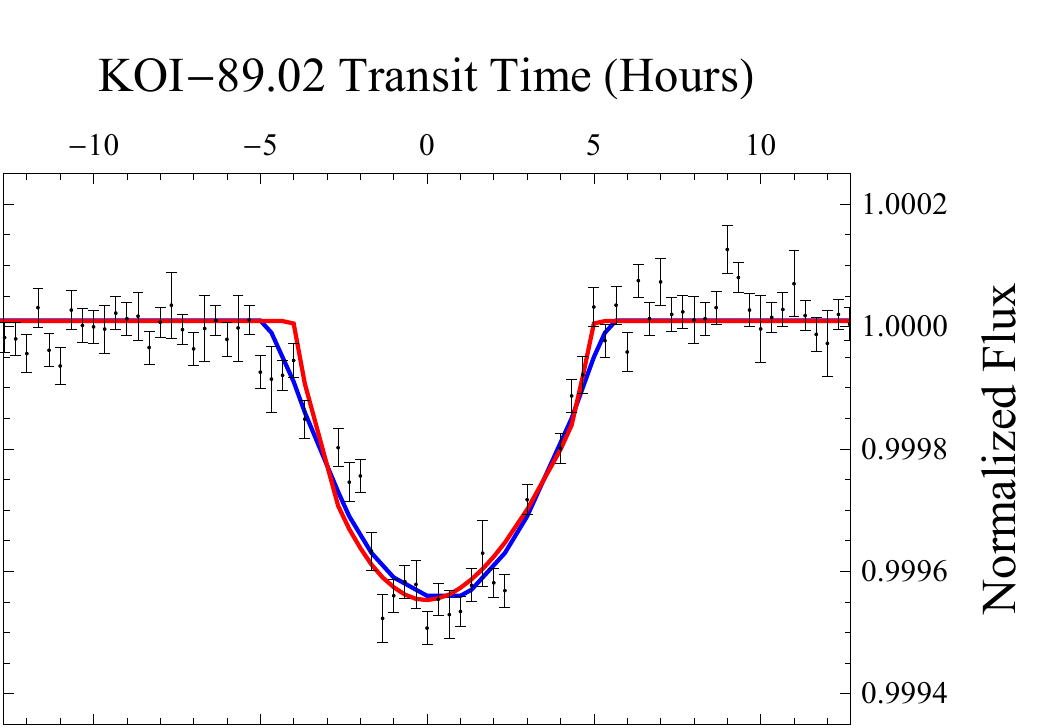}\\
\vspace*{-0.51cm}\\
\includegraphics[width=0.485692\textwidth]{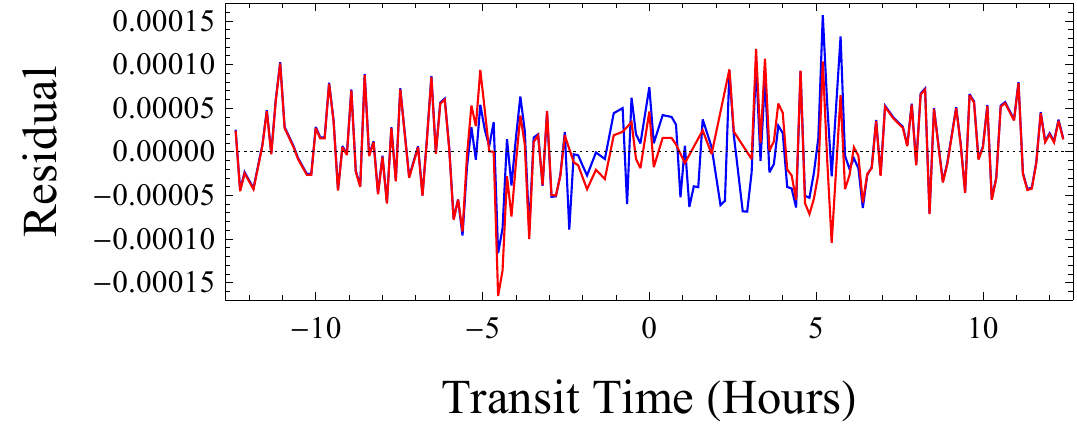} & \hspace*{-0.37cm}\includegraphics[width=0.485692\textwidth]{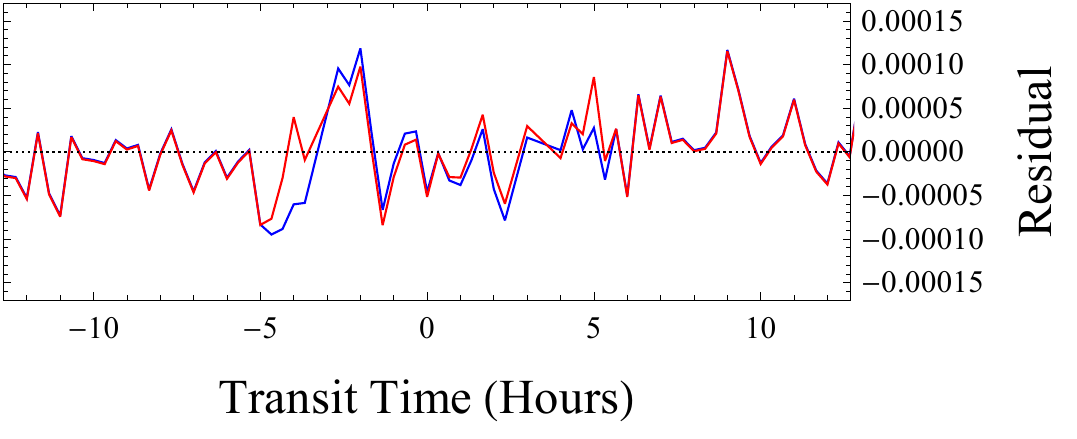} 
\end{tabular}
\caption{\footnotesize Best-fits and residuals of the KOI-89.01 and KOI-89.02 lightcurves. Red represents the gravity-darkened model, and blue represents grazing transits in the spherical model. The KOI-89.02 data are significantly noisier because of KOI-89.02's much longer orbital period, resulting in fewer total transits. We fit the two lightcurves simultaneously, resulting in a single best-fit line. The two lightcurves were placed side-by-side for visual comparison.}
\label{fig:fit}
\end{figure*}

The lightcurves of KOI-89.01 and KOI-89.02 (Figure \ref{fig:fit}) display unusual shapes. KOI-89.01 has the asymmetry expected of a misaligned body orbiting a fast-rotator \citep[J.][]{2009ApJ...705..683B} . KOI-89.02 does not display this asymmetry, possibly due to lower photometric precision. Both transits show sloped ingresses/egresses and entirely non-constant transit depths, producing dominant V-shaped lightcurves. A typical lightcurve is symmetric with a steep ingress and egress with a relatively flat bottom, rounded only by limb darkening.

KOI-89's V-shaped lightcurves can arise in one of two ways. First, planets only grazing their star during transit rather than fully eclipsing it block constantly changing sky-projected areas. This effect creates sloped ingresses/egresses. However, this situation is improbable for KOI-89 as both planets would require similar, high impact parameter values despite having significantly different semi-major axes. 

The second way KOI-89 could generate V-shaped lightcurve geometries is by having a gravity-darkened star with a very high stellar obliquity $\psi$ -- i.e. pole-on. In this case, the planets transit near a stellar pole and the gravity-darkened equator surrounds the outer edge of the star. The limb-darkening and gravity-darkening effects combine together to create a significant center-to-edge luminosity gradient. At ingress, the planet blocks a continuously increasing total flux as it moves closer toward the center of the star, and vice versa during egress. This produces a V-shaped transit lightcurve for each planet (J. \citet{2009ApJ...705..683B}, Figure 4), consistent with the lack of the typical ingress-egress asymmetry expected in a misaligned gravity-darkened transit. We test for the possibility of grazing transits by fitting the system with impact parameters nearly at and slightly above 1.0. We find that we can not match the system's lightcurve with grazing transits: such an event can not reproduce the proper ingress-egress asymmetry seen in KOI-89.01. 

Figure \ref{fig:fit} shows our best-fit lightcurve using grazing transits in blue. We apply grazing transits to both spherical and gravity-darkened models. We hold the stellar obliquity at zero in the gravity-darkened model to test the system for possible spin-orbit alignment. The poor fit of $\chi^2_{\mathrm{reduced}}=1.94$ (adjusted to account for holding the stellar obliquity constant) motivates us to investigate a model with a high stellar obliquity and rapid stellar rotation. 

Using the Levenberg-Marqhardt $\chi^2$ minimization technique, we fit for thirteen parameters:

\begin{itemize}
\item The stellar equatorial radius ($R_\star$)
\item The stellar obliquity ($\psi$)
\item The stellar normalized flux ($F_0$)
\item The radii of KOI-89.01 and KOI-89.02 ($R_{p_1}$, $R_{p_2}$)
\item The inclinations of KOI-89.01 and KOI-89.02 ($i_1$, $i_2$)
\item The sky-projected alignments ($\lambda_1$, $\lambda_2$)
\item The two orbits' eccentricities ($e_1$ , $e_2$) 
\item The center-of-transit times ($T_{0_1}$, $T_{0_2}$)
\end{itemize}

We display the best-fit lightcurve of our gravity-darkened model in Figure \ref{fig:fit} as the red line. 

\renewcommand{\arraystretch}{1.3}
\begin{table}[hbtp]
\centering
\begin{tabular}{l r}
\hline
{\bf Parameter} & {\bf Best Fit Values} \\ \hline
\hline
$\chi^2_\mathrm{reduced}$ & 1.52 \\
$R_{\star}$ & \rstar \\
$\psi$ & \obliq \\
$c_1$ (fixed) & \limbdarkone \\ 
$c_2$ (fixed) & \limbdarktwo \\
$\beta$ (fixed) & 0.25 \\ 
$F_0$ & \flux \\ 
$P_{\mathrm{rot}}$ (derived) & \rotperiod \\ 
$f_{\star}$ (derived) & \oblate \\
$R_\mathrm{p_1}$ & \rpone \\ 
$R_\mathrm{p_2}$ & \rptwo \\ 
$e_1 \geq$ & \eone \\
$e_2 \geq$ & \etwo \\
$i_1$ & \incone \\
$i_2$ & \inctwo \\ 
$b_1$ (derived) & \bone \\
$b_2$ (derived) & \btwo \\ 
$T_{0_1}$ & \tcenterone \\ 
$T_{0_2}$ & \tcentertwo \\
$\lambda_1$ & \ascone \\
$\lambda_2$ & \asctwo \\ 
$\varphi_1$ (derived) & \alignone \\ 
$\varphi_2$ (derived) & \aligntwo \\ 
 \\ 
\end{tabular}
\caption{\footnotesize Best-fit results for the KOI-89 system. We calculated stellar period of rotation $P_{rot}$ from $v\sin(i)$, $R_\star$, $M_\star$, and $\psi$. We derived the stellar oblateness $f$ from the Darwin-Radau relation. The impact parameters $b_1$ and $b_2$ were found using $P_1$ and $P_2$, $i_1$ and $i_2$, and $R_\star$. We set our limb-darkening parameters $c_1=u_1+u_2$ and $c_2=u_1-u_2$ according to \citet{sing2010stellar}.
}
\label{table:bestfit}
\end{table}

\subsection{Testing For TTV Systematics}

We test the $TTV$ ephemeris reported in \citet{rowe2014validation} for systematic errors by fitting the KOI-89 series as two epochs. The first epoch is comprised of KOI-89.01's first seven transits and KOI-89.02's first two transits. The second epoch is comprised of KOI-89.01's remaining six transits and KOI-89.02's remaining three transits. For each half, we adjust all transits with respect to their $TTV$ and fold the transits in the same fashion as described in \S \ref{sec:preparation}.

We apply our gravity-darkening model to both epochs and find that the resulting parameters of each fitted dataset have overlapping $1\sigma$ values with our best-fit values using the full timeseries (Table \ref{table:bestfit}). We therefore detect no evidence of systematics in the $TTV$ ephemeris listed in \citet{rowe2014validation}.

\subsection{Testing Limb Darkening Assumptions}
Limb-darkening has traditionally been problematic for the the gravity-darkening technique. Via Doppler Tomography, \citet{johnson2014misaligned} measured the sky-projected alignment of KOI-13.01 and found it to differ significantly from the gravity-darkening measurement performed in J. \citet{2011ApJS..197...10B}. \citet{masuda2015spin} proposed a solution to this discrepancy by demonstrating that the gravity-darkening model produces concurring measurements with Doppler Tomography when using a nonzero second quadratic limb-darkening term ($c_2$). This limb-darkening term is also a possible explanation for KOI-368's different spin-orbit misalignment values measured in  \citet{zhou2013highly} \citet{ahlers2014}. These works motivated us to update our gravity-darkening model to include both quadratic terms, $c_1$ and $c_2$.

KOI-89's very high stellar obliquity brings about an additional challenge in resolving limb-darkening. With the stellar pole near the center of the sky-projected stellar disk, the gravity-darkening and limb-darkening luminosity gradients behave nearly identically and are essentially additive. The resulting combined effects on a transit lightcurve are therefore degenerate with a stellar obliquity near $90^\circ$.

KOI-89's spectroscopically-determined effective temperature of $7717 \pm 225~\mathrm{K}$ corresponds to an approximate range of $0.55$ to $0.57$ for $c_1$ and $-0.165$ to $-0.135$ for $c_2$ \citep{sing2010stellar}. We test the robustness of our assumed limb-darkening values by refitting using $(0.55,-0.135)$ and $(0.57,-0.165)$ for $c_1$ and $c_2$, respectively. 

Applying the assumed $(c_1,c_2)$ values $(0.55,-0.135)$, we measure a slight increase in KOI-89.02's impact parameter; however, this increase vanishes when adjusting the gravity-darkening value $\beta$ to match Altair's value of $0.19$ \citep{monnier2007imaging}. We detect no significant changes in our best-fit results when employing the limb-darkening values $(0.57,-0.165)$. We cannot resolve the accuracy of our limb-darkening parameters without higher precision data, and therefore elect to apply the assumed values of $\beta=0.25$ \citep{von1924radiative} and $(c_1,c_2)$ values of $(0.56,-0.15)$  \citep{sing2010stellar}.

\subsection{Eccentricities} \label{sec:eccentric}

We constrain the lower limits of eccentricity to \eone~and \etwo, respectively. We address the degeneracy between eccentricity and argument of periapsis following \citet{price2015low}. Our eccentricities do not vary significantly for $|\omega|\leq150^{\circ}$ away from center-of-transit, consistent with \citet{price2015low} and J. \citet{barnes2007effects}. To find the lower limit for eccentricity, we set the center-of-transit at periapsis for both planets fit for the eccentricities using our gravity-darkened model.we analyze the plausibility of our eccentricity values in \S \ref{sec:eccstability}.

\subsection{Spin-Orbit Alignment} \label{sec:align}

Our gravity-darkened model results in a degeneracy in the sky-projected alignments between the values $\lambda$ and $180-\lambda$ \citep{ahlers2014}. We assume a prograde orbit for KOI-89.01, constraining $\lambda$ to a single value. This allows us to produce single, nondegenerate values for the obliquity ($\psi$) and each planet's inclination ($i$) in our best-fit model, which we allow to float in the full range of $0^\circ-360^\circ$.

We find that KOI-89 is highly misaligned with a stellar obliquity $\psi$ of \obliq, inclinations $i_1$ and $i_2$ of \incone~and \inctwo~respectively, and sky-projected alignments $\lambda_1$ and $\lambda_2$ of \ascone~and \asctwo~respectively. The high uncertainty of $\lambda_2$ is due to the apparent lack of asymmetry in KOI-89.02's lightcurve because of its photometrically imprecise data. We show our constraint of $\lambda_2$ in \S \ref{sec:stability}, which removes prograde/retrograde degeneracy via dynamic stability tests. 

With these constraints we calculate the true spin-orbit misalignments $\varphi_1$ and $\varphi_2$ from the equation \citep{winn2007transit},
\begin{equation}
\cos(\varphi_i)=\sin(\psi)\cos(i_i)+\cos(\psi)\sin(i_i)\cos(\lambda_i)
\label{eqn:align}
\end{equation}

modified for our parameter definitions. We calculate spin-orbit alignment angles of \alignone~and \aligntwo~for the two planets respectively. Table \ref{table:bestfit} lists all of KOI-89's parameter constraints.

\subsection{Double Transit} \label{sec:transit}

\begin{figure}[tbhp]
\epsscale{1}
\plotone{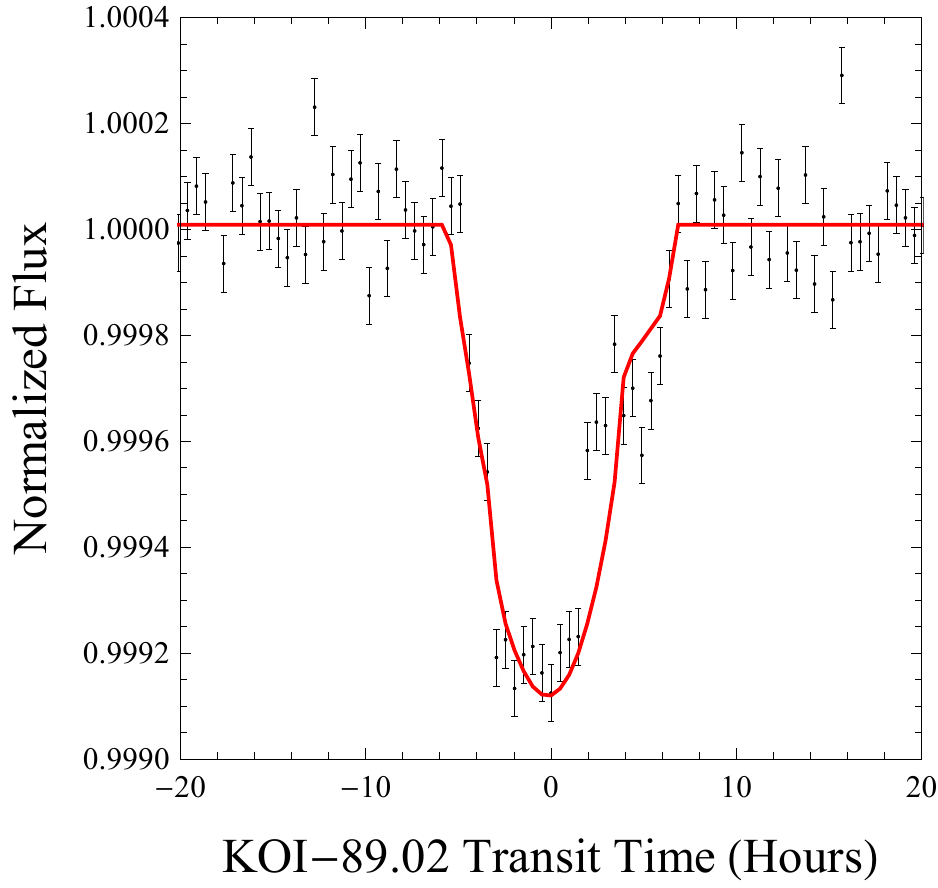}
\caption{\footnotesize Lightcurve of both planets transiting simultaneously. During the double transit, the depth is roughly double of a typical transit. Our best-fit model (in red) correctly reproduces the time of this event and the general shape of the lightcurve it produces. 
}
\label{fig:double}
\end{figure}

KOI-89.01 and KOI-89.02 simultaneously transit halfway through 2011, causing a significantly larger transit depth. We indicate this event with the arrow in Figure \ref{fig:alldata} and show the double transit and its synthetic lightcurve in Figure \ref{fig:double}. Our best-fit parameters produce a synthetic lightcurve that adequately models this event. We do not find evidence of a mutual event in the \emph{Kepler} dataset. 

\subsection{Oblateness} \label{sec:oblate}

We test the feasibility of KOI-89's oblateness value of \oblate~by analyzing its breakup rotation period,

\begin{equation}
P_{rot}=2\pi\sqrt{\frac{R_{\star}^{3}}{GM_{\star}}}
\end{equation}

We find that the star is rotating at $55\%-76\%$ of its breakup speed by calculating its rotation period, listed in Table \ref{table:bestfit}:
\begin{equation}
P_\star=\frac{2\pi R_\star\cos(\psi)}{v\sin(i)}
\end{equation}

This explains KOI-89's highly oblate shape and its gravity-darkened gradient, which are discussed in \S5.3. This star's oblateness is comparable to the fast-rotator Achernar with oblateness $\sim0.36$  \citep{carciofi2008determination} or other well-known oblate stars such as Altair ($\sim0.2$) \citep{monnier2007imaging}. Hence, our model produces physically plausible stellar parameters. 

\section{Dynamic Stability} \label{sec:stability}
\subsection{Coalignment Integrations}
Equation \ref{eqn:align} gives a planet's spin-orbit alignment $\varphi_i$ dependence on the sky-projected alignment $\lambda_i$. We fit for KOI-89's $\lambda_i$ in our gravity-darkening model, but are unable to resolve $\lambda_2$ due to its low photometric resolution and the host star's high stellar obliquity $\psi$. With gravity-darkening-driven asymmetry absent in KOI-89.02's lightcurve, we could not fully constrain its transit geometry. 

To estimate KOI-89.02's sky-projected alignment, we tested the system for dynamic stability for various transit geometries. Using our gravity darkening model, we constrained KOI-89's orbital elements with various assumed  $\lambda_2$ values. We then used the orbit integrator {\tt Mercury} from \cite{chambers1999hybrid} to test each orbit geometry for dynamic stability.

Using {\tt Mercury}, we perform mixed-variable symplectic (MVS) integrations of KOI-89 over $10^8$ years using 0.5 day timesteps. Using a spherical star allows for physically sound integrations that obey the conservation of angular momentum with minimal sacrifice; the stellar $J2\sim10^{-4}$ value (calculated following \citet{ssdynamics}), coupled with the planet's long orbit periods, cause nodal precession on a timescale that would not significantly affect the system's stability. We assume ice-giant densities of $\rho=1.64~\mathrm{g/cm^3}$ for both planets. 

We define an angle $\alpha$ of coalignment between the two orbits, defined relative to their angular momentum vectors:
\begin{equation}
\alpha \equiv \cos^{-1}\left(\frac{\vec{L}_1 \cdot \vec{L}_2}{|\vec{L}_1||\vec{L}_2|}\right)
\end{equation}

\begin{figure}[tbhp]
\epsscale{1.1}
\plotone{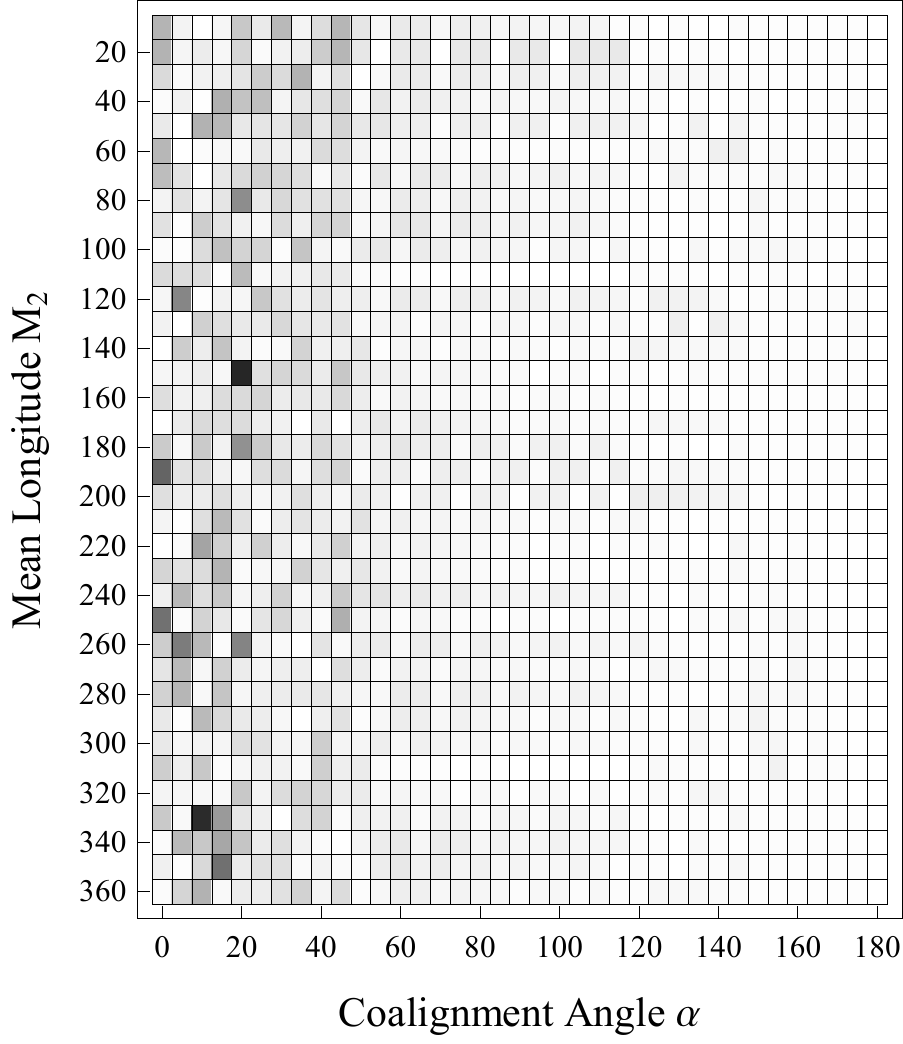}
\caption{\footnotesize Survival times for various initial configurations of the KOI-89 system. Darker color indicates longer longer survival time, with the longest survival time $5.1\times10^7$ years. Beyond $\alpha=20^\circ$, ejection/collision events occur very quickly for all initial configurations, suggesting that the KOI-89 system is more stable near coalignment. 
}
\label{fig:survival}
\end{figure}

Figure \ref{fig:survival} shows the survival time of KOI-89 as a function of the coalignment angle $\alpha$ and conjuction longitude. By varying KOI-89.02's longitude of periapsis, we vary the conjuction longitude between the two planets. We define system instability as a planet ejection or collision event. None of our 360 simulations produced a stable orbit for $10^8$ years, indicating we have not found a physically viable system yet. In general, survival times are longer for lower $\alpha$, but the longest-lived architectures are non-planar. 

%We find that that this system is only stable with low planet densities ($\rho\lesssim1.64~\mathrm{g/cm^3}$). KOI-89.01 and KOI-89.02 are likely ice or gas giants, consistent with the planets' measured radii of \rpone~and \rptwo~\citep{swift2012mass}.  

The coalignment angle $\alpha$ is approximately the difference between the two planets' sky-projected alignment angles. Using the results of our orbital integrations, we estimate the difference between the sky-projected alignment angles $\alpha\approx|\lambda_2 - \lambda_1|$ to be $20^\circ \pm 20^\circ$. This is a conservative estimate based on our results in Figure \ref{fig:survival}; follow-up observations would provide a much better calculation of this parameter.

Our 1332 orbit integrations resulted in a maximum survival time of $5.1\times10^7$ years. The lack of stable configurations suggests that this system is not yet fully understood. If the system is in resonance and is non-planar, it may evolve chaotically \citep[R.][]{barnes2014anti}, and hence long-lived configurations may only exist in small ``islands" of parameter space. Alternatively, KOI-89's stability could be brought about by unknown additional bodies in the system. We  show in \S \ref{sec:eccstability} that KOI-89 could be stable if KOI-89.02's eccentricity is lower than our best-fit value of \etwo.   A better characterization of this system's stability could be understood via $TTV$ analysis or Rossiter-McLaughlin measurements, but such work is outside the scope of this project.

\subsection{Eccentric Integrations} \label{sec:eccstability}
In addition to our coalignment/mean longitude stability tests, we also test the stability of KOI-89.02's eccentricity of \etwo~in a coplanar configuration. \citet{2015ApJ...808..126V} demonstrated that, in general, multiplanet systems have low eccentricities, making KOI-89 a potential exception to the rule. See \S \ref{sec:eccentric} for an explanation  of our treatment of longitude of periapsis.

\begin{figure}[tbhp]
\epsscale{1.0}
\plotone{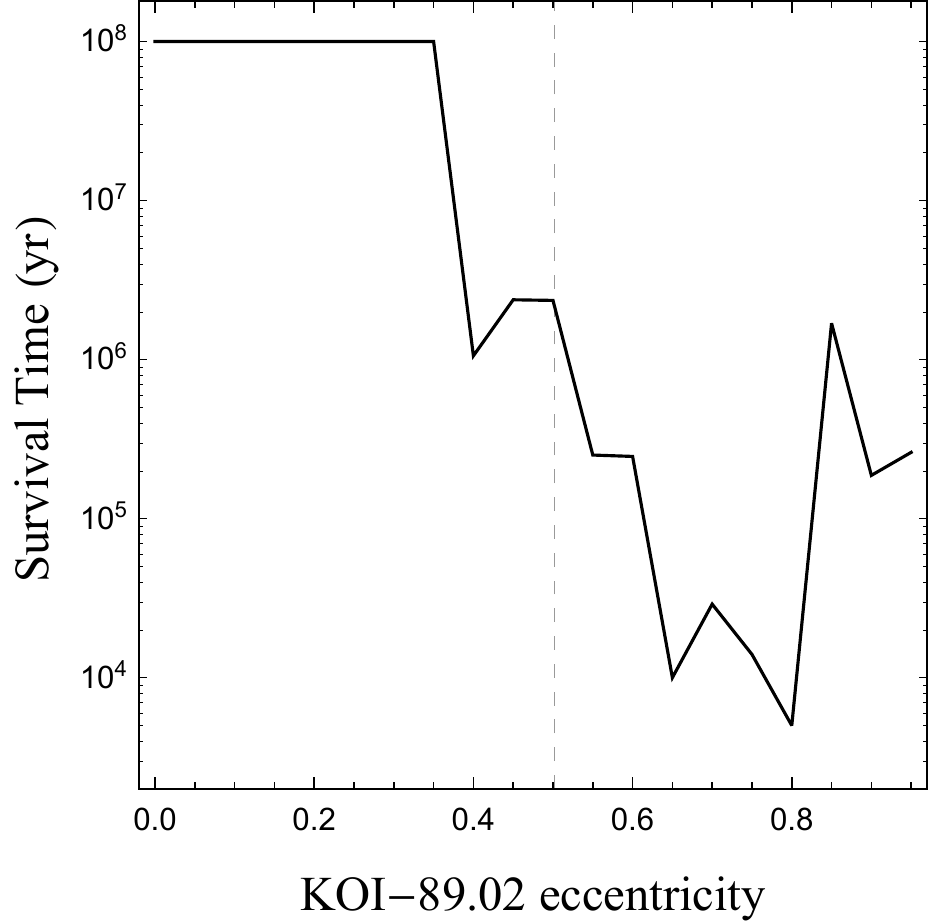}
\caption{\footnotesize KOI-89 survival times using various assumed $e_2$ values in a coplanar configuration. Our N-body integrations are stable through $10^8$ years for all $e_2 \leq 0.35$, which is less than two $\sigma$ of our best-fit value. The dashed line marks our best-fit value of \etwo.
}
\label{fig:eccentricity}
\end{figure}

We perform a series of integrations in {\tt Mercury} \citep{chambers1999hybrid} using assumed $e_2$ values ranging from 0.0 to 0.95 and a 0.05 step size. All $e_2 \leq 0.35$ are stable, roughly consistent with \citet{2015ApJ...808..120P}. We show the results of these integrations in Figure \ref{fig:eccentricity}.

Not surprisingly, lower $e_2$ values yield longer lifetimes and overall higher stability. This result suggests three possibilities. The first (and least likely) possibility is that this system is in fact not coplanar. If the fitted $e_2$ value of \etwo~is correct, then perhaps higher stabilities are found in slightly non-coplanar orbits. While higher stability in such a configuration is counterintuitive, it does at least  reduce the odds of a close encounter between the two planets, limiting the chances of a violent collision/ejection event.

The second possibility is that our eccentricity measurement contains systematics. A grazing transit would reduce the transit duration time similarly to an eccentric orbit transiting near periapsis, and could produce a V-shaped lightcurve like we see in Figure \ref{fig:fit}. KOI-89.02's low signal-to-noise ratio, coupled with the degeneracy between impact parameter and planet radius that arises in all grazing transits, prevents us from resolving whether KOI-89.02 is in fact fully eclipsing its host star. KOI-89.02's TTVs could also drive up our eccentricity measurement if they are not fully accounted for \citep{2015ApJ...808..126V}. High-precision follow-up photometry could better determine KOI-89.02's orbit parameters, including its eccentricity.

The third possibility is that unknown bodies in the system provide stability to these orbits. \citet{2015arXiv150901131A} demonstrated that highly eccentric orbits in or near mean-motion resonance can exhibit long-term stability. Additional bodies could help stabilize KOI-89.01 and KOI-89.02, explaining why our best-fit parameters do not display dynamic stability through $10^8$ years in our orbit integrations.

%\subsection{Inclined Resonance} \label{sec:resonance}
%
%The KOI-89 system is nearly in a 5:2 resonance. When testing the system's dynamic stability, we checked for resonant arguments following Rory \citet{barnes2014anti}. We calculate the inclined resonant arguments in our dynamically stable {\tt Mercury} integrations and find evidence of a 5:2 resonance in several integrations. All stable systems display some pattern in the resonant argument, 
%\begin{equation}
%\theta=5\lambda_2-2\lambda_1-2\varpi_1-\varpi_2
%\end{equation}
%with a $\sim5000$ year timescale. Figure \ref{fig:resonant} shows an example of the above resonant argument over 30000 years.

%\begin{figure}[tbhp]
%\epsscale{1}
%\plotone{resonant}
%\caption{\footnotesize An example of the near-resonant behavior found in our orbit integrations. This integration had a starting coalignment angle of $\alpha=19.03^\circ$ and assumed planet densities of $\rho=1.64~\mathrm{g/cm^3} $. We found evidence of near-resonance at varying strengths in all orbit integrations surviving at least $10^6$~years.}
%\label{fig:resonant}
%\end{figure}

%KOI-89's near-resonant behavior could imply that the two planets' eccentricities and inclinations evolve chaotically \citep[Rory][]{barnes2014anti}. A more in-depth analysis of these phenomena would be possible with a better constraint of the two planets' argument of periapsis, and with a better understanding of KOI-89's $TTVs$. However, this supposed near-resonant behavior could be much more complicated if unknown bodies exist in this system. 

\section{Discussion} \label{sec:discussion}

\begin{table*}[tbpH]
\centering
\begin{tabular}{p{0.3\linewidth} p{0.325\linewidth} p{0.1\linewidth} p{0.1\linewidth} p{0.1\linewidth}}
\hline
{\bf Mechanism} & {\bf Relevant Works} & {\bf Criteron 1} & {\bf Criteron 2} & {\bf Criteron 3} \\ \hline
\hline

Star-disk-binary interactions & \citet{lai2014star}, \citet{xiang2014evolution}, \citet{batygin2012primordial} & Yes & Yes & Inconclusive \\

Inclination driven by a warped disk & \citet{terquem2013effects} & Yes & Yes & Inconclusive \\

Planet-planet scattering & \citet{chatterjee2008dynamical}, \citet{ford2005planet}, \citet{raymond2008mean}, \citet{nagasawa2008formation} & Yes & Yes & Inconclusive \\

Kozai resonance & \citet{libert2009kozai}, \citet{thies2011natural}, \citet{payne2010transit} \citet{2011ApJ...742L..24K} & Yes & Yes & Inconclusive \\

Internal Gravity Waves & \citet{2012ApJ...758L...6R}, \citet{2015arXiv150207779F} & Yes & Yes & Inconclusive \\

Planet-embryo collisions & \citet{levison1998modeling}, \citet{charnoz2001short} & Yes & No & No \\ 

Chaotic evolution of stellar spin & \citet{storch2014chaotic}, \citet{valsecchi2014tidal} & No & Yes & Yes \\ 

Magnetic torquing & \citet{lai2011evolution}, \citet{spalding2014primordial} & No & Yes & Yes \\

Coplanar high-eccentricity migration & \citet{petrovich2014hot} & Yes & No & Yes \\

Inclination Resonance & R. \citet{barnes2014anti} & No & Yes & Inconclusive \\

\end{tabular}
\caption{\footnotesize Possible spin-orbit misalignment mechanisms for the KOI-89 system. We list \nomech possible causes of spin-orbit misalignment that have been put forward in the literature and rule out \noimposs of them based on our best-fit results and our estimation of the two planets' coalignment. The three criteria are: (1) consistency with KOI-89's fundamental parameters, (2) the capability to cause extreme misalignment, and (3) the production of mutually aligned planets.}
\label{table:mechanisms}
\end{table*}

The KOI-89 system is highly misaligned with spin-orbit alignment angles of \alignone~and \aligntwo~for the two planets respectively. Our preliminary dynamical analysis of the system \S\ref{sec:stability} failed to find a stable solution, so at this time we cannot rigidly constrain the mutual inclination. However, we recognize that survival times are longer in general for lower $\alpha$.

Of \nomech misalignment mechanisms suggested in the literature, \noposs are consistent with
our results. We rule out the other \noimposs mechanisms based on three criteria:
\begin{enumerate}
\item Consistency with KOI-89's fundamental parameters such as orbit period, stellar radius, etc.  
\item The capability to cause extreme misalignment
\item Conformance with mutually aligned planets
\end{enumerate}
We compare our results to each mechanism in Table \ref{table:mechanisms}.

\subsection{Star-Disk-Binary Interactions}
\citet{batygin2012primordial} first showed that a stellar companion could warp a star's protoplanetary disk into misalignment. Planets could then form in the plane of the disk, resulting in primordial spin-orbit misalignment \citep{lai2014star, xiang2014evolution, batygin2012primordial}. This mechanism requires an unknown binary star in the KOI-89 system, but fundamentally agrees with our results in that it could produce highly misaligned, coplanar orbits.   

\subsection{Inclination Driven By A Warped Disk}
A planet in the potential of a warped protoplanetary disk can be driven to very high misalignment values \citep{terquem2013effects}. \citet{teyssandier2012orbital} found that Jupiter-mass planets misaligned from a warped disk experience dynamic friction that realigns the planet in timescales shorter than the lifetime of the disk. However, Neptune-mass planets can remain misaligned and have their eccentricities driven up by orbital perturbations from the disk's gravitational potential. This mechanism has only been applied to single-planet systems, so criterion 3 is inconclusive. However, this mechanism agrees with the first two criteria and cannot be ruled out based on our results.

\subsection{Planet-Planet Scattering}
Our results cannot entirely rule out planet-planet scattering, which is orbit migration due to close encounters between high-mass objects \citep[e.g.][]{chatterjee2008dynamical, ford2005planet, raymond2008mean, nagasawa2008formation}. With KOI-89's net orbital angular momentum highly misaligned from the star's spin angular momentum, conservation of angular momentum would require additional planet(s) to scatter the known two planets. In this scenario it is highly unlikely that the two planets would end up near mutual alignment. However, if a sufficiently large unknown body exists in this system, then our orbit integrations are unsound and our coalignment constraint for this system is invalid. We therefore deem this mechanism consistent as a possible cause of KOI-89's misalignment. Further studies of KOI-89's $TTVs$ could confirm the existence of additional planets.

\subsection{Kozai Resonance}
Kozai resonance in the KOI-89 system requires an unknown body that is significantly misaligned with its known orbital plane \citep{libert2009kozai,thies2011natural,payne2010transit}. Such an event would likely not produce coplanar orbits for KOI-89.01 and KOI-89.02. However, \citet{2011ApJ...742L..24K} suggests that coplanar, inclined orbits might arise as a result of this mechanism. We therefore deem this method consistent.

\subsection{Internal Gravity Waves}
\citet{2012ApJ...758L...6R} showed that angular momentum transport between the convective interior and radiative exterior of hot, early-type stars can change the observed stellar spin axis, resulting in spin-orbit misalignment. This misalignment mechanism happens independently of orbiting bodies and does not affect coplanarity. The 2-D simulations performed in \citet{2012ApJ...758L...6R} found that this mechanism can occur on a timescale as short as tens of years, and can explain retrograde orbits. Whether this mechanism can produce spin-orbit misalignments near $90^\circ$ is still under investigation.

\subsection{Planet-Embryo Collisions}
Planet-embryo collisions can occur in any standard formation model, and they can drive migration in various ways \citep{levison1998modeling, charnoz2001short}. However, this mechanism can produce large spin-orbit misalignment angles only for small rocky bodies and does not apply to the highly misaligned giant planets KOI-89.01 and KOI-89.02. Additionally, this mechanism likely could not produce coplanar misaligned orbits because the collisions driving this mechanism are unique to each planet.

\subsection{Chaotic Evolution of Stellar Spin}
\citet{storch2014chaotic} and \citet{valsecchi2014tidal} demonstrated that strong tidal dissipation can cause chaotic evolution of stellar spin. This mechanism requires hot Jupiters with periods $\lesssim 3~\mathrm{days}$. Such a body in the KOI-89 system would have to be drastically misaligned from the plane of the other two orbits; therefore, this mechanism cannot be the standalone cause of misalignment because some other mechanism would have to misalign the hot Jupiter. If there was a non-transiting hot Jupiter that was initially misaligned, it could torque the star into misalignment with the other planets. If said hot Jupiter fell into its host star because of tidal decay, it could change both KOI-89's rotation axis and rotation rate \citep{jackson2009observational}. However, early-type stars such as KOI-89 have weak tidal interactions in general \citep{ogilvie2007tidal}.

\subsection{Magnetic Torquing}
Magnetic torquing between a stellar magnetic field and a protoplanetary disk can cause misalignment by torquing the disk away from the star's equatorial plane
\citep{lai2011evolution,spalding2014primordial}. KOI-89 is an early-type star with a weak magnetic field \citep{bagnulo2002measuring}, so this mechanism could not cause KOI-89's high misalignment. We note that the magnetic fields of fast-rotators are still under investigation \citep{ibanez2015stability}; a better understanding of these magnetic fields may reveal this to be a possible misalignment mechanism for KOI-89.

\subsection{Coplanar High-Eccentricity Migration}
Coplanar high-eccentricity migration can occur in mutually aligned multiplanet systems with at least one highly eccentric orbit. Secular gravitational effects excite the inner planet's eccentricity to very high values, and planetary tidal dissipation during periapsis reduces the orbit's semi-major axis. This mechanism occurs primarily in the planets' orbital plane, predominantly maintaining the system's original spin-orbit alignment angles \citep{petrovich2014hot}.

\subsection{Inclination Resonance}
If the planets are in resonance and possess a mutual inclination, then the orbital inclinations can be driven to very large values \citep[R.][]{barnes2014anti}. In that case we may expect to find at least one planet in a misaligned orbit. This phenomenon can also produce very large eccentricities. However, this mechanism depends on stellar torquing from tidal interactions for both planets to be discovered in a misaligned state. Such tidal interaction is weak around early-type stars \citep{ogilvie2007tidal}. While this is a possible cause of KOI-89's extreme misalignment, it requires an external mechanism to bring about an initial mutual inclination. More work is needed to understand if this scenario is possible and could apply to KOI-89.

\section{Conclusion}

We constrain the individual spin-orbit alignments of multiplanet system KOI-89. With our gravity-darkened model, we found significant spin-orbit misalignment with angles of \alignone~and \aligntwo~for KOI-89.01 and KOI-89.02, respectively. We also constrain other fundamental parameters of the KOI-89 system and estimate the mutual alignment between KOI-89.01 and KOI-89.02. We show these results in Table \ref{table:bestfit}.

While our measurements alone do not uniquely assign a misalignment mechanism to KOI-89, the large spin-orbit alignment angles $\varphi_i$ and low coalignment angle $\alpha$ of this system limit the possible mechanisms for planet migration.  These values, the measured $TTVs$, the near 5:2 resonance, and the fast rotation of the star itself all imply a dynamic formation history. 
  
KOI-89 is of of particular interest because it can experimentally constrain the numerous outstanding hypotheses that have been proposed to generate misalignment. We limit possible causes to star-disk-binary interactions, disk warping via planet-disk interactions, planet-planet scattering, or internal gravity waves in the convective interior of the star. Follow-up observations searching for additional bodies could provide evidence for any of these hypotheses, including internal gravity waves if no additional bodies are found.

Much could still be learned about the KOI-89 system. Asteroseismic determination of the star's oscillation modes could confirm various stellar properties such as stellar radius, mass, and obliquity. Doppler tomographic observations could constrain the individual ascending nodes of the two planets. Analysis of the $TTV$ could confirm/deny the existence of undiscovered planets in the system. High-precision photometry could better constrain the two planets' eccentricities and impact parameters and help resolve the degeneracy between limb-darkening and gravity darkening. Any of these follow-up observations would shed new light on the formation of solar systems dissimilar to our own. 

The constraints provided in this work add to the sample of known misaligned systems -- particularly misaligned multiplanet systems, of which only a small number are currently known. The unique nature of the KOI-89 system provides new insight for studying system formation and evolution. It also adds to the surprising diversity of exosystems discovered to date. Future studies can apply the knowledge gained from this work to a wide variety of misaligned and dynamic systems.

We would like to thank the \emph{Kepler} Science Team for making this work possible -- particularly Dr. David Latham for providing the $v\sin(i)$ measurement of this system. JPA and JWB are funded by NASA Proposal \#13-ADAP13-213. RB acknowledges support from NSF grant AST-1108882.

\bibliographystyle{apj}
\bibliography{citations}

\end{document}